\documentclass[11pt,a4paper]{article}
\pdfoutput=1
\usepackage{jheppub}
\usepackage{color}
\usepackage{multirow}

\begin{document}

\title{Excited and exotic bottomonium spectroscopy from lattice QCD}

\author[a]{Sin\'{e}ad~M.~Ryan,} \emailAdd{ryan@maths.tcd.ie}
\author[b]{David~J.~Wilson} \emailAdd{d.j.wilson@damtp.cam.ac.uk}

\affiliation[a]{School of Mathematics and Hamilton Mathematics Institute, Trinity College, Dublin 2, Ireland}
\affiliation[b]{DAMTP, University of Cambridge, Centre for Mathematical Sciences, Wilberforce Road, Cambridge, CB3 0WA, UK}

\affiliation{\vspace{0.1cm} {\rmfamily \normalsize (for the Hadron Spectrum Collaboration)}}

\abstract{We explore the spectrum of excited and exotic bottomonia using lattice QCD. Highly excited states are identified with masses up to 11,000~MeV, many of which can be grouped into supermultiplets matching those of the quark model 
while exotic spin--parity--charge-conjugation quantum numbers $J^{PC}=0^{+-},\,1^{-+},\,2^{+-}$ that cannot be formed from $\bar{q}q$ alone are also identified. Single-meson operator constructions are used that have good $J^{PC}$ in the continuum, these are found to overlap well onto heavy quark states with $J\le4$. A continuum $J^{PC}$ is assigned to each level, based on the distribution amongst lattice irreps and dominant operator overlaps. States with a dominant gluonic component are identified and form a hybrid supermultiplet with $J^{PC}=(0,1,2)^{-+},\, 1^{--}$, approximately 
1500~MeV above the ground-state $\eta_b$, similar to previous computations with light, strange and charm quark systems.}
\maketitle

\section{Introduction}
\label{sec:intro}

The spectrum of bottomonium mesons has been extensively studied using 
effective field theories, sum rules, models and ab initio
theoretical methods. More recently, the discovery of unexpected, narrow 
charmonium-like states - the so-called XYZs - has led to a resurgence of
interest in heavy quark spectroscopy
fuelling theoretical discussions as to their nature and structure. 
Suggestions include
hydrid mesons, molecular or tetraquark configurations or simple
quark-antiquark mesons albeit with unexpected masses. In addition, in the
plethora of newly discovered states no observation of a state with manifestly exotic quantum
numbers, that cannot be produced by quark-antiquark constructions alone, has been made. To date an explanation of
the charmonium-like XYZs as well as the absence of charmonium and bottomonium 
states with exotic $J^{PC}$ has not emerged. 

A similar abundance of new states has yet to emerge in the bottomonium sector, where the meson
spectrum remains relatively poorly explored, although this will be addressed in the near future by LHCb and
Belle II~\cite{Bediaga:2018lhg,Kou_2019}. A recent review~\cite{Brambilla:2019esw} summarises 
the current theoretical and experimental results in both charm and bottom-quark systems. 
Quark models appear to account for many of the low-lying states, although many are yet to be 
found. Nevertheless,
five hadronic states containing a $b\bar{b}$ component, which are inconsistent with quark models
have already been discovered. In a similar pattern to the charmonium XYZs, three of these states have
conventional quantum numbers $\Upsilon(10580,10860,11020)$ while two states, the (charged)
$Z_b(10610)$ and
$Z_b(10650)$ appear to require a four-quark configuration, for example in a compact tetraquark or a molecular
arrangement. Calculations of the spectrum of excited and exotic bottomonium mesons are therefore timely.

Relatively little is known about the presence of exotic hadrons in bottomonium from lattice QCD. However, exotic hadrons have been predicted in both integer and half-integer spins, in a series of studies considering light, strange~\cite{Dudek:2009qf,Dudek:2010wm,Dudek:2011tt,Edwards:2011jj,Dudek:2011bn,Dudek:2012ag}, and charm quarks~\cite{Liu:2012ze,Moir:2013yfa,Padmanath:2013zfa,Padmanath:2015jea,Cheung:2016bym}. Evidence has also emerged from exploratory lattice QCD studies of compact tetraquark constructions containing $b$-quarks, that some channels may admit bound-state solutions~\cite{Francis:2016hui,Francis:2018jyb,Junnarkar:2018twb,Leskovec:2019ioa}. The large separation of scale between bottom and light quarks makes reliable first-principles lattice QCD computations challenging. Demanding that discretisation effects of ${\cal O}(am_Q)$ for both light and bottom quarks are small, while simultaneously ensuring that the volume is large enough 
($m_\pi L > 4$) requires very large and very fine lattices. 

Effective field theories computed on the lattice, including NRQCD and the Fermilab approach, have played an important role in bridging the gap between maintaining a reasonable computational cost and accessing the physics of interest. In recent work heavy quarks have also been simulated with 
a relativistic highly-improved staggered quark action (heavy-HISQ) and a 
range of heavy quark masses are used to enable a reliable extrapolation to the bottom quark mass.   
An impressive array of precision $b$-physics quantities has been determined and the methods themselves are now proven to be robust and reliable. It is relatively commonplace to extract ground-state quantities precisely, and several studies have also reported success in computing excited state spectra~\cite{Meinel:2009rd,Meinel:2010pv,Dowdall:2011wh,McNeile:2012qf,Aoki:2012xaa,Lewis:2012ir,Dowdall:2013jqa,Wurtz:2015mqa,Bailey:2017nzm}. 

In this study we present results from an exploratory relativistic dynamical anisotropic 
lattice calculation of the excited spectrum of bottomonium. On an anisotropic lattice the 
lattice spacing in the temporal direction, $a_t$, is finer than in the spatial direction, $a_s$ 
providing enhanced resolution for specroscopy. Earlier work using anisotropic lattices for 
heavy quark physics is described in Refs.~\cite{Liao:2001yh,Foley:2004jf}. 
This work follows Refs.~\cite{Liu:2012ze,Cheung:2016bym} where charm quarks were simulated using a relativistic Wilson-clover quark action (the same action as used for the light quarks) on an anisotropic lattice with stout-smeared~\cite{Morningstar:2003gk}
spatial links, and distillation for quark smearing~\cite{Peardon:2009gh}. We apply this relativistic action to bottom quarks for the first time, 
noting that while $a_sm_b>1$, in the temporal direction $a_tm_b$ remains less than unity.
The effect of spatial discretisation is investigated through dispersion relations and is 
found to be mild, at least within the scope of this calculation. 

In this work we demonstrate that the techniques developed and used in Refs.~\cite{Dudek:2009qf,Dudek:2010wm,Dudek:2011tt,Edwards:2011jj,Dudek:2011bn,Dudek:2012ag,Liu:2012ze,Moir:2013yfa,Padmanath:2013zfa,Padmanath:2015jea,Cheung:2016bym} for spin 
identification from large correlation matrices formed of large bases of operators and analysed variationally, allow for the 
identification of extensive spectra of bottomonia, including quark-model-like 
and hybrid supermultiplets spanning conventional 
$q\bar{q}$ and exotic quantum numbers. Some preliminary results from these calculations 
were presented in Ref.~\cite{Ryan:2020spd}. 

The paper is organised as follows: Sections~\ref{sec:calc_details} and~\ref{sec:fits_spins} 
describe the details and parameters of this lattice calculation. The bottomonium spectrum is presented in 
Section~\ref{sec:results}, where the determination exotic states and a 
supermultiplet of hybrid mesons is also discussed. A summary and outlook is presented in 
Section~\ref{sec:summary}.

\section{Calculation Details}
\label{sec:calc_details}

We use $2+1$ flavours of dynamical quarks on an anisotropic lattice  with a temporal lattice spacing, $a_t$ that is finer than the spatial lattice spacing $a_s$, with anisotropy $\xi= a_s/a_t\approx 3.5$. The light quarks are tuned to produce a heavier-than-physical pion of approximately 391 MeV. A tree-level Symanzik-improved action is used
in the gauge sector while the fermion sector is described by a tadpole-improved Sheikholeslami-Wohlert (clover) action, including stout-smeared spatial gauge
fields~\cite{Morningstar:2003gk} and distillation quark smearing~\cite{Peardon:2009gh}. Full details of the action and parameters can be found in Refs.~\cite{Edwards:2008ja,Lin:2008pr}.

Table~\ref{tab:lattice_details} summarises the lattice ensembles used in this study. The final 
results are produced using two time-sources on an $(L/a_s)^3\times T/a_t=20^3\times128$ volume at a single value of the lattice spacings ($a_s,a_t$). To test the volume dependence, a single time-source on a $24^3$ volume is used. Dispersion relations are extracted from 
energy levels at finite momentum computed with a single time-source on the $20^3$ volume.
\begin{table}[h]
  \begin{center}
    \begin{tabular}{c|c|c|c|c}
      Lattice Volume & $m_\pi$ (MeV) & $N_{\rm cfgs.}$ & $N_{\rm tsrcs}$ & $N_{\rm vecs.}$ \\
      \hline
      $20^3\times 128$ & $391$ & 603 & 1-2 & 128\\
      \hline
      $24^3\times 128$ & $391$ & 553 & 1 & 162\\
    \end{tabular}
    \caption{Details of the lattice gauge field ensembles used in this study. The volume is
      given as $(L/a_s)^3\times (T/a_t)$ where $L$ and $T$ are the spatial and
      temporal extents of the lattice. The number of gauge field configurations, the
      number of (perambulator) time-sources per configuration and the number of eigenvectors
      employed in the distillation method are given as $N_{\rm cfgs.}, N_{\rm tsrcs}$ and
      $N_{\rm vecs.}$ respectively.}
      \label{tab:lattice_details}
  \end{center}
\end{table}

\subsection{Lattice heavy quarks}
\label{subsec:hq}
In an anisotropic lattice computation with dynamical quarks, the bare gauge and fermion 
anisotropies,
which enter in the Monte Carlo ensemble generation, must be simultaneously
tuned to produce a target value when measured non-perturbatively~\cite{Edwards:2008ja}. 
For heavy valence quarks, an additional tuning of the fermion anisotropy 
is useful to ensure a consistent measured anisotropy~\cite{Liu:2012ze}. The tuning condition is the same as for
light quarks namely that the speed of light, $c=1$, or in other words that a
relativistic dispersion relation is measured. For charm quarks, it was
 demonstrated that a single valence anisotropy parameter, tuned with the pseudoscalar
($\eta_c$) dispersion relation, also yields a relativistic
dispersion relation for the vector ($J/\psi$) and for the heavy-light ($D$-meson) sector, consistent within uncertainties.

The dispersion relation also plays an important role in the Fermilab treatment
of lattice heavy quarks~\cite{ElKhadra:1997}. In this approach a mass-dependent tuning
of parameters guarantees a smooth interpolation between the light and heavy
sectors. The bottom quark mass in the simulation is tuned so that the
kinetic mass measured from the quarkonium dispersion relation takes its
physical value and need not be the same as the rest mass. 
In a relativistic simulation the rest and kinetic masses are equal but this is not required
in the Fermilab approach, while in non-relativistic formulations of QCD (NRQCD) the rest mass is discarded.

In this work the bottom quark is treated relativistically, computed with the same tadpole-improved Wilson-clover action used for the light quarks. The bottom quark mass parameter in the simulation is determined by imposing that the pseudoscalar meson mass takes the physical $\eta_b$ value~\cite{Tanabashi:2018oca}. We investigate the occurrence and significance
of discretisation errors using dispersion relations, and mass splittings of the pseudoscalar and 
vector ground states. 
The relativistic form of the dispersion relation for a meson $A$ can be written 
\begin{equation}
  (a_t E_A)^2 = (a_t m_A)^2 + \left(\frac{1}{\xi_A}\right)^2 (a_s p)^2, 
  \label{eqn:disp}
\end{equation}
and since the quark fields in the meson satisfy periodic boundary conditions for the finite
extent of the lattice, the momenta are quantised as 
$a_s \vec{p} = \frac{2\pi}{L}(n_x,n_y,n_z)$ with $n_i\in\mathbb{Z}$ and 
for simplicity we label momenta by $[n_xn_yn_z]$ in the text.  
The renormalised anisotropy, $\xi_A$
can be determined from the slope of the dispersion relation.
The input anisotropy parameter in the valence heavy-quark action was tuned so that the
pseudoscalar ($\eta_b$) dispersion relation produces $\xi\approx 3.5$. Keeping the heavy-quark mass and the input
anisotropy fixed, the anisotropy determined from the heavy-heavy vector ($\Upsilon$) and in heavy-light ($B$-meson) systems was then compared with the $\eta_b$. 

\begin{figure}[h]
  \begin{center}
    \includegraphics[width=0.99\textwidth]{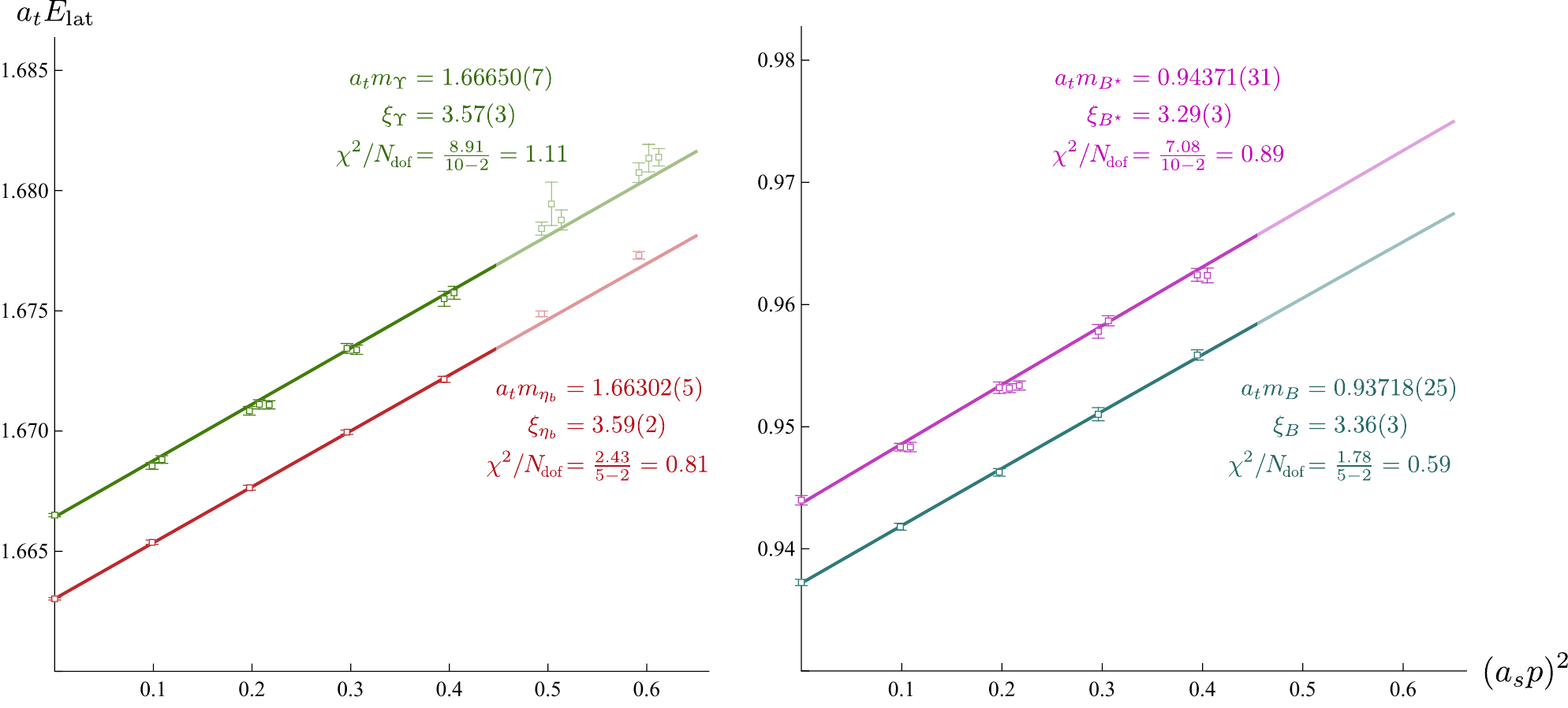}
    \caption{Dispersion relations for $\eta_b$, $\Upsilon$, $B$ and $B^\star$ hadrons. 
In the case of vector states where multiple helicity components are present, $|\lambda|=0$ is 
the leftmost point, the $|\lambda|=1$ points to the right of this have been displaced slightly 
from their true $(a_sp)^2$ values for legibility. 
The faded points in the left panel are produced by momentum $[210]$ and $[211]$ operators. 
The corresponding energy levels are less reliably determined and are not included in 
the fits shown. The line shown corresponds to fits with $n^2=n_x^2+n_y^2+n_z^2\le 4$ in 
Table~\ref{tab:disp} and the errors on the points and the fit parameters are statistical only.}
    \label{fig:combined_disps}
  \end{center}
\end{figure}

The dispersion relations for the $\eta_b$ and $\Upsilon$ mesons are shown in the left panel of Figure~\ref{fig:combined_disps}, for a range of momenta up to $\vec{p}=[211]$\;\footnote{Details regarding the construction of moving frame hadron operators are given in Ref.~\cite{Thomas:2011rh}.}. 
At each momenta a large basis of interpolating operators has been used in a variational 
analysis to determine the energy level in the appropriate 
lattice irreducible representation, as described in sec.~\ref{sec:fits_spins}. 
The rest mass and anisotropy determined from fits to Eq.~\ref{eqn:disp} are summarised in Table~\ref{tab:disp}.
The goodness-of-fit ($\chi^2/N_\mathrm{dof}$) values obtained show the data are well-described 
by Eq.~\ref{eqn:disp}. Little evidence of deviation from the relativistic form of the dispersion 
relation is apparent below momentum $[210]$, while [210] and [211] appear to show a small 
differences. We note that for these latter high momenta points 
the correlation functions only plateau at large euclidean times and the energy levels are less 
robustly determined. 
The measured anisotropies are $\xi_{\eta_c}=3.590(15)$ and $\xi_\Upsilon=3.574(26)$, within 3\% and 2\% of the target value 
respectively. 

To probe the reliability of the simulations further we determine the pseudoscalar and vector dispersion
relations in the $B$-meson sector using the same input parameters as for the $\eta_b$ and
$\Upsilon$. The measured anisotropy is determined from the slope of the dispersion relation and comparing this
parameter in the heavy-heavy and heavy-light sectors effectively isolates momentum-dependent discretisation effects
providing a clean test of the robustness of heavy-quark simulations. This was highlighted in an analysis of differences in binding energies between heavy-heavy and heavy-light systems in Refs.~\cite{Collins:1995qp,Kronfeld:1996uy}. 
The dispersion relations for $B$-mesons are shown in the right-panel of Figure~\ref{fig:combined_disps}. 
In this case, the measured anisotropies of the pseudoscalar and vector heavy-light mesons 
are $\xi_B=3.360(29)$ and $\xi_{B^\ast}=3.293(31)$, within 4\% and 6\% of the target value respectively. 
As noted, there is some discrepancy between the anisotropy measured in the bottomonium and $B$ meson systems but 
in each case the measured values are at most 6\% from the target anisotropy. 
A similar, although smaller, effect was observed between charmonium and $D$ mesons. Improved determinations of the energy 
levels at finite momentum and tuning on a larger volume may reduce this discrepancy, which will be investigated in future work. 
In conclusion, the dispersion relations for both heavy-heavy and heavy-light vector and pseudoscalar mesons yield precise 
determinations of the measured anisotropy demonstrating relativistic behaviour including for large values of
spatial momentum. 

\begin{table}
\begin{center}
\begin{tabular}{c|ccccc}
           & $a_t m$     & $\xi$      & $n^2\le$ & $\chi^2/N_\mathrm{dof}$\\
\hline
\multirow{2}{*}{$\eta_b$}
           & 1.66302(5)  & 3.590(15)  & 4                              & $\frac{2.44}{5  - 2} = 0.81$ \\
           & 1.66305(5)  & 3.568(12)  & 6                              & $\frac{9.04}{7  - 2} = 1.81$ \\
\hline
\multirow{2}{*}{$\Upsilon$} 
           & 1.66650(7)  & 3.574(26)  & 4                              & $\frac{8.91}{10  - 2} = 1.11$ \\
           & 1.66650(9)  & 3.534(20)  & 6                              & $\frac{18.99}{16  - 2} = 1.36$ \\
\hline
$B$        & 0.93718(25) & 3.360(29)  & 4                              & $\frac{1.78}{ 5  - 2} = 0.59$ \\
\hline
$B^\star$  & 0.94371(31) & 3.293(31)  & 4                              & $\frac{7.08}{10  - 2} = 0.89$ 
\end{tabular}
\caption{Parameters determined from dispersion relation fits using Eq.~\ref{eqn:disp} to rest and moving-frame energies, in $J^P=0^-,1^-$ for both heavy-heavy and heavy-light systems, $n$ is defined below Eq.~\ref{eqn:disp}. The energies used and the fit for $n^2\le 4$ are shown in Fig.~\ref{fig:combined_disps}.}
\label{tab:disp}
\end{center}
\end{table}

\subsection{Spatial discretisation effects and the hyperfine splitting}
\label{subsec:hfs}
The hyperfine splitting inferred from the masses in Table~\ref{tab:disp}\;\footnote{To quote 
energies in physical units we use the ratio of the mass of the $\Omega$ baryon measured in a 
lattice calculation~\cite{Edwards:2011jj} to its experimental mass, yielding $a_t^{-1}=5667$ MeV.}
is significantly smaller than the experimental value. A similar effect was noted using this approach in charmonia, as reported in Ref.~\cite{Liu:2012ze}. This underestimate is a common feature of calculations at finite lattice spacing. Nevertheless, 
several other approaches have determined the hyperfine splitting using lattice QCD~\cite{Meinel:2009rd,Meinel:2010pv,Dowdall:2011wh,Aoki:2012xaa,Dowdall:2013jqa,Bailey:2017nzm}, in many cases finding good agreement. 

The (spatial) clover term $c_s$ appears in Eq.~5 of Ref.~\cite{Edwards:2008ja}, and its tree-level tadpole-improved value is used consistently for the dynamical quarks and the valence bottom quark computed here. However, this is expected to underestimate the non-perturbative value~\cite{Luscher:1996ug} required for full ${\cal O}(a)$ improvement. The effect of increasing $c_s$ was explored for charmonia in Ref.~\cite{Liu:2012ze} where it was found that by replacing the tree level value $c_s=1.35$ with $c_s=2$ the $1S$ hyperfine splitting increased from $\Delta_{c_s=1.35}=80(2)$ MeV to $\Delta_{c_s=2.0}=114(2)$ MeV, significantly closer to the physical value of 117 MeV.

A similar analysis is repeated here for bottomonium and summarised in Table~\ref{tab:csw}. A clover coefficient $c_s=2.2$ was chosen to approximate the
nonperturbative value at the bottom quark mass, and the $\eta_b$--$\Upsilon$ mass difference was determined. 
Empirically, the hyperfine splitting increases with increasing $c_s$,
$\Delta_{c_s=1.2}=19.5\pm 0.6$ MeV to $\Delta_{c_s=2.2}=55.3\pm 0.4$ MeV, much closer to the experimental value of 61~MeV.
Disconnected contributions are not included in this analysis, as was also the case in charmonium, however these are
expected to be small due to OZI suppression and are unlikely to play a significant role in the discrepancies with experimental values discussed above. 
\begin{table}
\begin{center}
\begin{tabular}{c|ccc||c}
$c_s$                          & 1.2          & 2.0              & 2.2           & experiment\\
\hline
$(m_\Upsilon-m_{\eta_b})/$~MeV & 19.5$\pm$0.6 & 47.4$\pm$0.8     & 55.3$\pm$0.4  & 61.6$\pm$2.0
\end{tabular}
\caption{A comparison of the the $1S$ hyperfine splitting determined
      with the tree-level clover coefficient and with increased values,
      $c_s=2.0$ and $2.2$ to investigate the possible effects of nonperturbative
      improvement and to
      provide an approximate scale for the discretisation error}
\label{tab:csw}
\end{center}
\end{table}

We conclude that for the anisotropic action with stout smearing used here, the ${\cal O}(a_sp)$ discretisation errors are
under reasonable control for bottomonium and heavy-light simulations at zero and finite momenta up to $[200]$.
The latter is important as in subsequent work we will extend our analysis to include meson-meson operators for states above decay threshold. Finally, the 
analysis of the effect of enlarging the clover term suggests a scale for the 
discretisation errors of approximately 35-40 MeV.

\section{Operator construction, fitting and spin identification}
\label{sec:fits_spins}
Following the notation and methods developed and described in Refs.~\cite{Dudek:2009qf,Dudek:2010wm,Liu:2012ze},
meson energies are determined from two-point correlation functions
\begin{equation}
  {\cal C}_{ij}(t) = \langle 0|{\cal O}_i(t){\cal O}_j^\dagger (0)|\rangle ,
\end{equation}
where the operators ${\cal O}_j^\dagger (0)$ create the state of interest at $t=0$ which is later
annhilated by ${\cal O}_i(t)$ at euclidean time $t$. The correlation functions
have a spectral decomposition
\begin{equation}
  {\cal C}_{ij}(t) = \sum_{\mathfrak{n}}\frac{Z_i^{\mathfrak{n}\ast}Z_j^{\mathfrak{n}}}{2E_{\mathfrak{n}}}e^{-E_\mathfrak{n} t} ,
\end{equation}
where $Z_i^{\mathfrak{n}} =\langle\mathfrak{n}|{\cal O}_i^\dagger|0\rangle$ are the operator overlaps. Distillation is used to create correlation functions,
facilitating the efficient construction of large operator bases. A derivative-based construction is used, combining gauge-covariant
forward-backward, spatial derivatives and gamma matrices in a fermion bilinear to form operators of
definite $J^{PC}$. The discrete lattice breaks rotational symmetry, and the relevant symmetry group for a lattice calculation is $O_h$ whose five, ten including parity, irreducible representations (irreps) are labelled $A_1, T_1, T_2,E, A_2$. The distribution of a
continuum spin state across the lattice irreps is shown in Table~\ref{tab:irreps}, for $J\leq 4$. The
assignment of continuum spin values to lattice energy levels, characterised by lattice irreps, is
not straightforward, the procedure we follow is discussed in Refs.~\cite{Dudek:2009qf,Dudek:2010wm}.
Finally, the lattice operators used in this work are constructed to transform in a definite lattice irrep
and row, $\Lambda$ and $\lambda$ respectively, and are derived from continuum operators ${\cal O}^{J,M}$ by subduction.
\begin{equation}
  {\cal O}_{\Lambda, \lambda}^{[J]} = \sum_M {\cal S}_{J,M}^{\Lambda,\lambda}{\cal O}^{J,M},
  \label{eq:subduction}
\end{equation}
where $M$ is the $J_z$ component of spin. In each irrep we include operators in the basis that are 
proportional to the commutator of two covariant derivatives, the field-strength tensor, to investigate the pattern of hybrid
mesons. The number of operators in each lattice irrep, $\Lambda^{PC}$,
used in this study is given in Table~\ref{tab:irreps}.
\begin{table}[htb]
  \begin{center}
    \begin{tabular}{c|l}
      J & $\Lambda$(dimension) \\
      \hline
      0 & $A_1(1)$ \\
      1 & $T_1(3)$ \\
      2 & $T_2(3)\otimes E(2)$  \\
      3 & $T_1(3)\otimes T_2(3)\otimes A_2(1)$   \\
      4 & $A_1(1)\otimes T_1(3)\otimes T_2(3)\otimes E(2)$\\
    \end{tabular}
    \hspace{1cm}
        \begin{tabular}{c|cccc}
      $\Lambda$ &       $\Lambda^{-+}$ &      $\Lambda^{--}$ &   $\Lambda^{++}$ &  $\Lambda^{+-}$ \\
      \hline
      $A_1$     & 12    & 6  & 13 & 5  \\
      $T_1$     & 18    & 26 & 22 & 22 \\
      $T_2$     & 18    & 18 & 22 & 14 \\
      $E$       & 14    & 12 & 17 & 9 \\
      $A_2$     & 4     & 6  & 5  & 5 \\
        \end{tabular}
        \caption{The left table shows the distribution of continuum spin across the lattice 
irreps at rest, up to
          $J=4$. The right table lists the number of operators used in each lattice irrep. The operators are
          fermion-bilinears of the form
          $\bar{\psi}\Gamma D_i D_j\ldots \psi$ constructed with
          combinations of forward-backward derivatives and gamma matrices to have definite momentum,
          as discussed in the text.}
        \label{tab:irreps}
  \end{center}
\end{table}

\begin{figure}[!t]
  \begin{center}
    \includegraphics[width=0.99\textwidth]{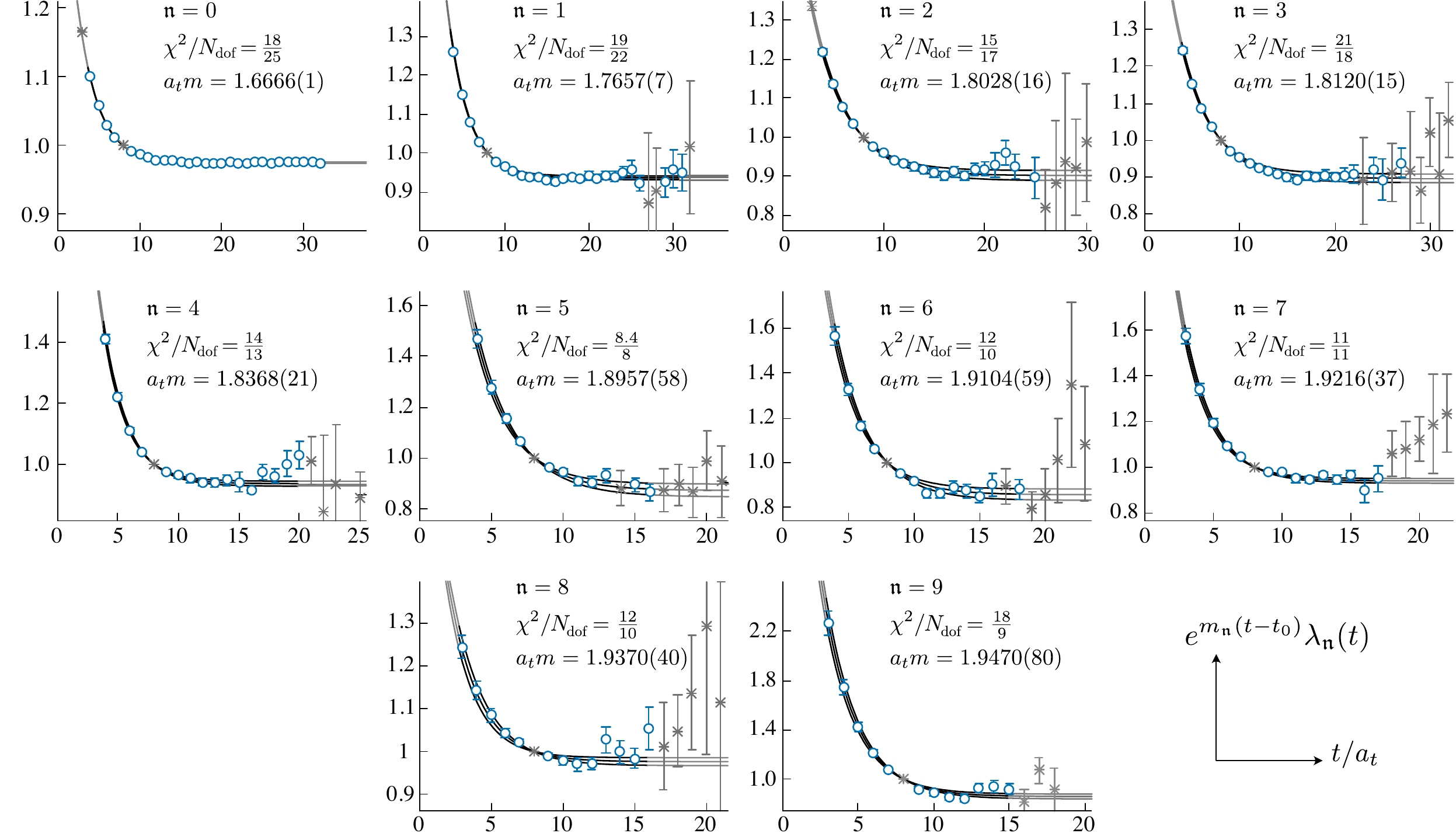}
    \caption{Principal correlator fits, according to Eq.~\ref{eqn:princorr}, in the $T_1^{--}$ irrep 
extracted on $t_0/a_t=8$ on the $(L/a_s)^3=20^3$ volume computed with two time-sources. 10 levels are identified. Data points 
are $\lambda^{\mathfrak{n}}(t)\cdot e^{m_{\mathfrak{n}(t-t_0)}}$ with fits shown including the one sigma statistical 
uncertainty. The grey stars were not included in the fit.}
    \label{fig:princorr}
  \end{center}
\end{figure}

The energy levels and operator overlaps are extracted from two-point correlation functions using the
variational method~\cite{Michael:1985ne,Luscher:1990ck}. In each lattice
irrep energies and overlaps are extracted from a matrix of two-point correlators, $C_{ij}(t)$ by solving a generalised eigenvalue problem
\begin{equation}
  C_{ij} (t) v^{\mathfrak{n}}_j  = \lambda^{\mathfrak{n}}(t,t_0) C_{ij} (t_0) v ^{\mathfrak{n}}_j ,
\end{equation}
where $i$ and $j$ label the operators and $t_0$ is the reference timeslice.
The generalised eigenvalues, or principal correlators, $\lambda^{\mathfrak{n}}$ yield the energies from fits to 
\begin{equation}
  \lambda_{\mathfrak{n}}(t) = (1-A_{\mathfrak{n}})e^{-m_{\mathfrak{n}}(t-t_0)} + A_{\mathfrak{n}}e^{-m^\prime_{\mathfrak{n}}(t-t_0)},
\label{eqn:princorr}
\end{equation}
with free parameters $m_{\mathfrak{n}}, m^\prime_{\mathfrak{n}}, A_{\mathfrak{n}}$, where $m_\mathfrak{n}$ is the energy of 
the state of interest. An example of the principal correlators determined in this study is shown in Figure~\ref{fig:princorr}. 
In general we find this method continues to work well with bottom quarks, as it also did in studies of the excited charmonium and 
open charm mesons and for light (isovector and isoscalar) mesons and baryons.

\subsection{Spin identification}
Lattice energy levels are assigned continuum spin values following the approach described in
Refs.~\cite{Dudek:2009qf,Dudek:2010wm}. The lattice operators used in this work are
subduced from continuum operators with definite spin, as detailed in Eq.~\ref{eq:subduction}, empirically it has been observed that they do not depend significantly on the specific lattice irrep into which they are subduced.
Continuum spin assignments are then made by comparing the overlap values determined independently
in different lattice irreps which have been subduced from the same continuum operator.
This is a powerful tool to discriminate the spin of higher-lying energy levels especially in a dense spectrum with
many states that are close to degenerate in mass. This method was shown to work well in the charmonium spectrum~\cite{Liu:2012ze} and the same methodology is
employed here.

\begin{figure}[!t]
\begin{center}
\includegraphics[width=0.5\textwidth]{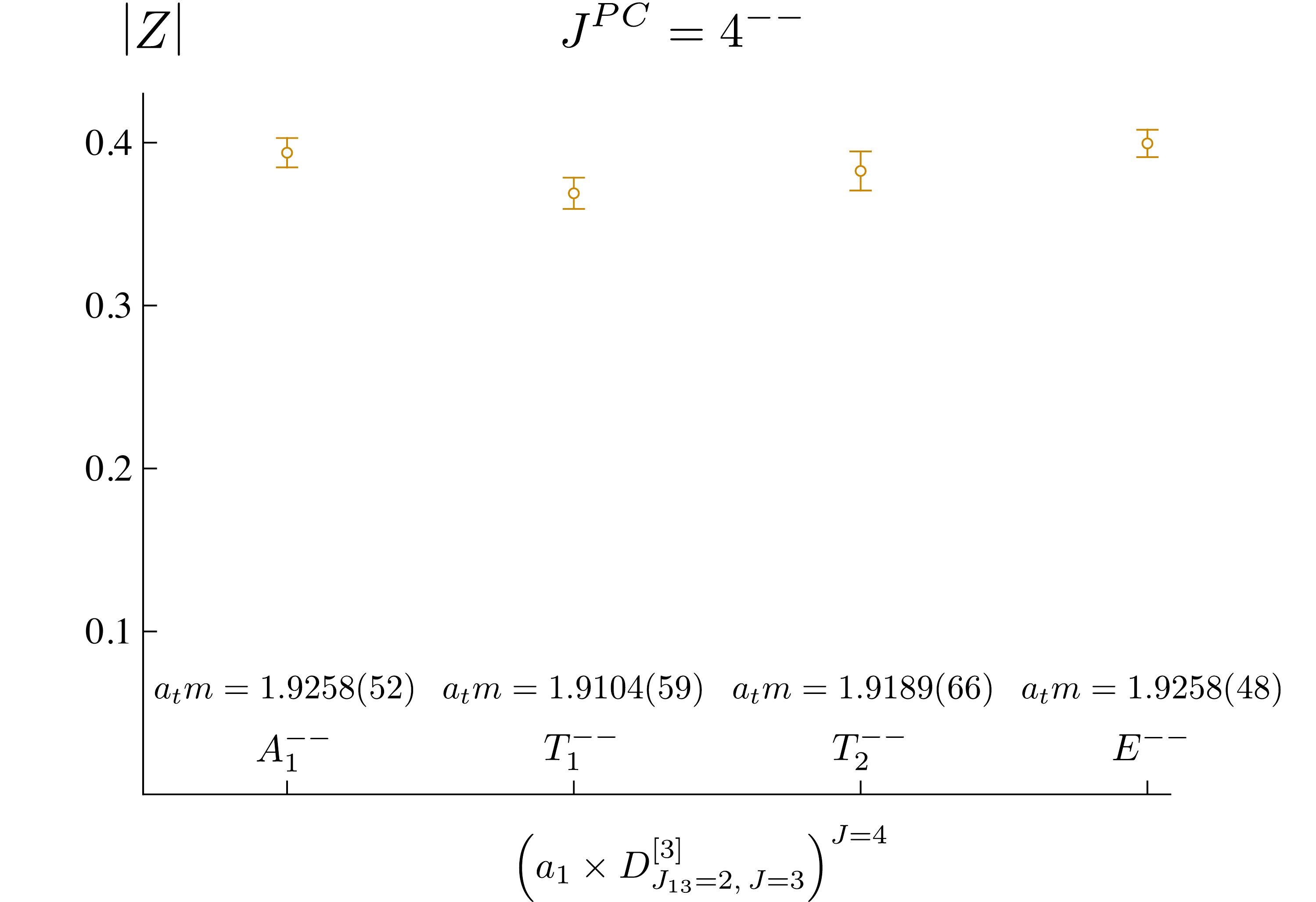}
\caption{Operator overlap values $|Z|$ for a continuum $J^{PC}=4^{--}$ operator subduced into lattice irreps $A_1^{--}$, $T_1^{--}$, $T_2^{--}$ and $E^{--}$. The masses and overlaps are found to be quite consistent across lattice irreps. We thus identify these four states as components of the same continuum $4^{--}$ level.}
\label{fig:overlaps_4mM}
\end{center}
\end{figure}

An illustrative example is given in Figure~\ref{fig:overlaps_4mM}, which shows the overlap values $Z^\mathfrak{n}_i$ determined in the variational analysis for the $\left(a_1\times D^{[3]}_{J_{13}=2,\,J=3}\right)^{J=4}$ operator\footnote{Operator definitions are provided in Ref.~\cite{Dudek:2010wm}.} with overall $J^{PC}=4^{--}$. This appears in $A_1^{--}$, $T_1^{--}$, $T_2^{--}$ and $E^{--}$ when subduced and we find the operator produces a consistent value across irreps. In general we find good agreement
between $Z$ values that originate from the same continuum operator, and we use this approach to assign continuum
spin values up to $J=4$.

\begin{figure}[h]
\begin{center}
\includegraphics[width=0.999\textwidth]{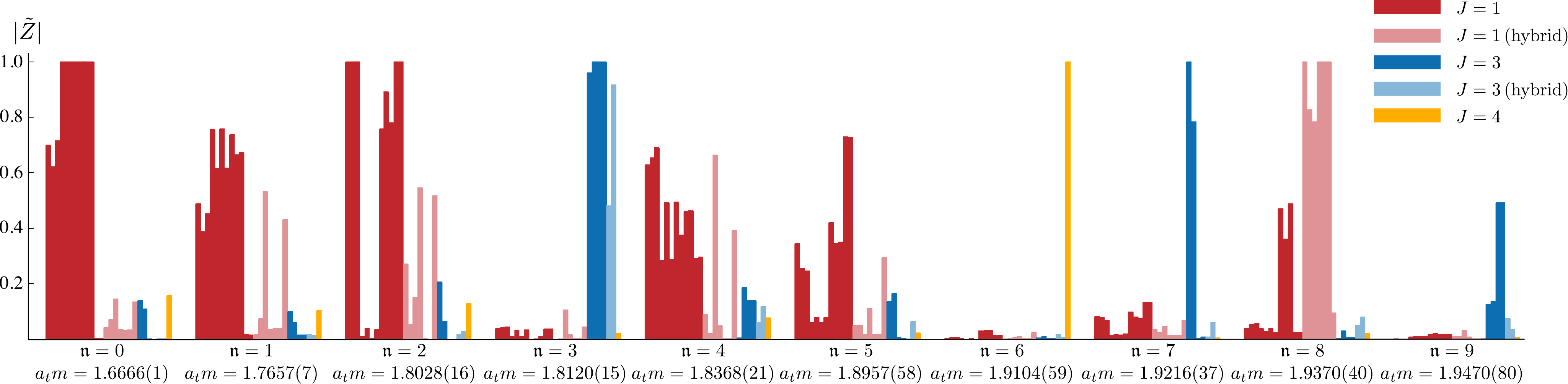}
\caption{Normalised operator overlaps $|\tilde{Z}|$ for the lowest 10 states in the $T_1^{--}$ irrep on the $20^3$ volume. The 
overlaps are normalised so that the largest value across all states is unity. Each state is labelled by its mass in lattice units determined in a variational analysis. The colours represent the continuum spin from which the operator was subduced and lighter colours indicate hybrid-like operators.}
\label{fig:skyscrapers}
\end{center}
\end{figure}

An illustrative example of the \emph{relative} operator overlaps for each state $\mathfrak{n}$ is given in Figure~\ref{fig:skyscrapers}, here the overlap for operator $i$ is plotted using $\tilde{Z^\mathfrak{n}_i}=Z^\mathfrak{n}_i/\mathrm{max}_{\mathfrak{\forall m}}(|Z^\mathfrak{m}_i|)$. Even in this densely packed $T_1^{--}$ irrep, the spin assignment is abundantly clear.

\section{Results}
\label{sec:results}
In this section we present results for the spectrum of bottomonium mesons
including candidate hybrid states. The results, organised by the relevant
lattice irrep are shown first, followed by the final spin-identified spectrum
labelled by the continuum quantum numbers $J^{PC}$. 

\subsection{The bottomonium spectrum by lattice irrep and volume comparison}
\label{subsec:spectrum_irreps}
\begin{figure}[h]
  \begin{center}
    \includegraphics[width=0.99\textwidth]{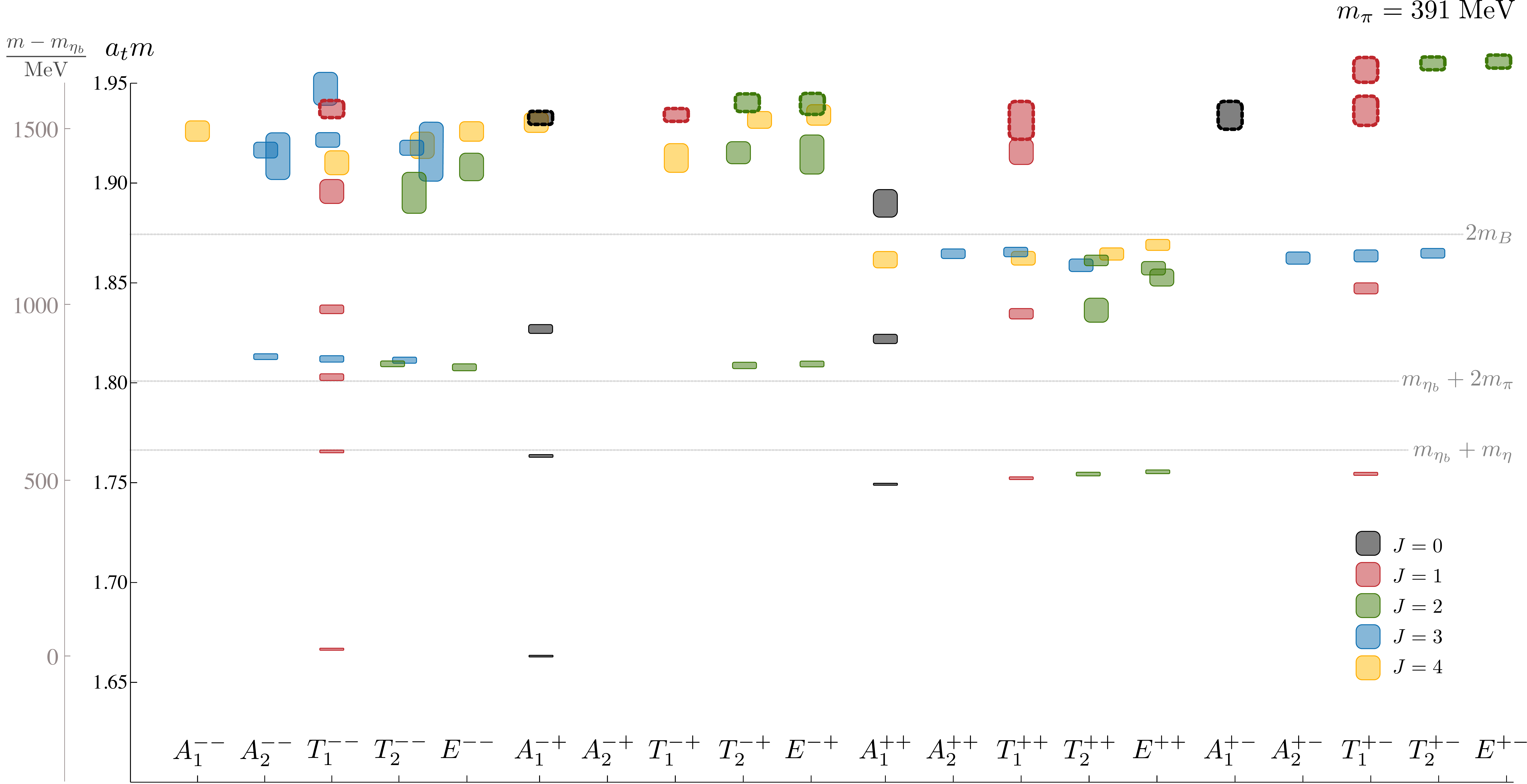} 
    \caption{The spectrum of bottomonium mesons determined on the $20^3$ volume with two time sources and organised by lattice irrep. The states are color coded according to the spin of the continuum operator which has the dominant overlap in each level. The height of the boxes is the one-sigma uncertainty about the mean. The dashed lines show the lowest relevant thresholds from masses determined in this calculation. Shown are the robustly determined energy levels in each irrep. States above $a_tE\approx 1.96$ are not shown.}
    \label{fig:irrep_spectrum}
  \end{center}
\end{figure}

The spectrum of excited and exotic bottomonium states determined on the $20^3$ volume and arranged by lattice irreps, labelled $\Lambda^{PC}$, 
is shown in Figure~\ref{fig:irrep_spectrum}. The results are presented relative to the energy of the lowest state in 
the system, the $\eta_b$ meson, to mitigate the uncertainty due to tuning heavy quark mass. 
The vertical height of the boxes shows the one sigma statistical uncertainty
about the mean, from a determination of energy levels using the variational analysis, as described earlier. The energy levels are colour-coded
according to the spin, identified as described above
(black for $J=0$, red for $J=1$, green for $J=2$, blue for $J=3$ and gold for
$J=4$), as in Ref.~\cite{Liu:2012ze}.

\begin{figure}[h]
  \begin{center}
    \includegraphics[width=0.99\textwidth]{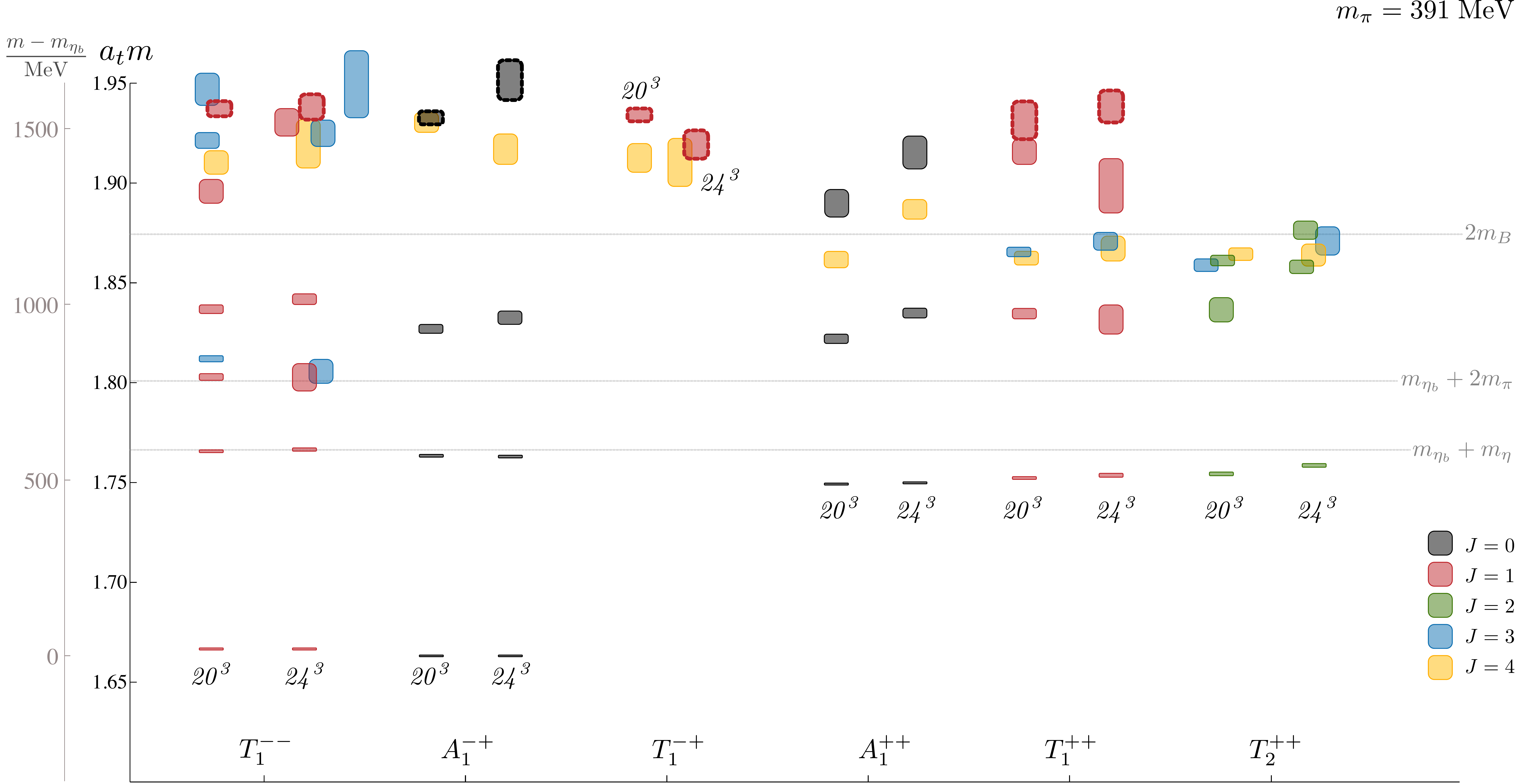} 
    \caption{A comparison between the spectra obtained in selected irreps at two spatial volumes $(L/a_s)^3=20^3,\,24^3$. }
    \label{fig:irrep_spectrum_L20_L24}
  \end{center}
\end{figure}

In Figure~\ref{fig:irrep_spectrum_L20_L24} the volume dependence of the extracted spectrum is investigated, 
by comparing results from two spatial volumes, $20^3$ and $24^3$. Overall the same pattern of states is observed although   
with some dependence on volume visible  particularly in the higher-lying states, 
albeit with larger statistically uncertainties on the $24^3$ volume. 

In this study single-meson operators constructed from fermion bilinears are utilised. 
Since no multi-hadron operators are included a strong overlap onto multi-hadron states is not expected. 
Investigations of the consequences of this choice have found that using only 
approximately-local operators, a level usually arises within the width of the state~\cite{Dudek:2012xn,Wilson:2015dqa}. 
We find that the calculated spectrum is well-described by single mesons as indicated by the only mild volume dependence 
and very consistent agreement for a state of definite continuum spin subduced into its relevant lattice irreps. 
However, we cannot rule out the presence of additional spectroscopic states that may be found by extracting and 
analysing the scattering amplitude~\cite{Dudek:2016cru,Briceno:2017qmb,Hansen:2019nir} and we note that relevant 
two-hadron strong thresholds include $\eta_b\eta$, $\eta_b\pi\pi$ and $B\bar{B}$. 
\subsection{The spin-identified bottomonium spectrum}
\label{subsec:spectrum_final}
\begin{figure}[h]
  \begin{center}
    \includegraphics[width=0.99\textwidth]{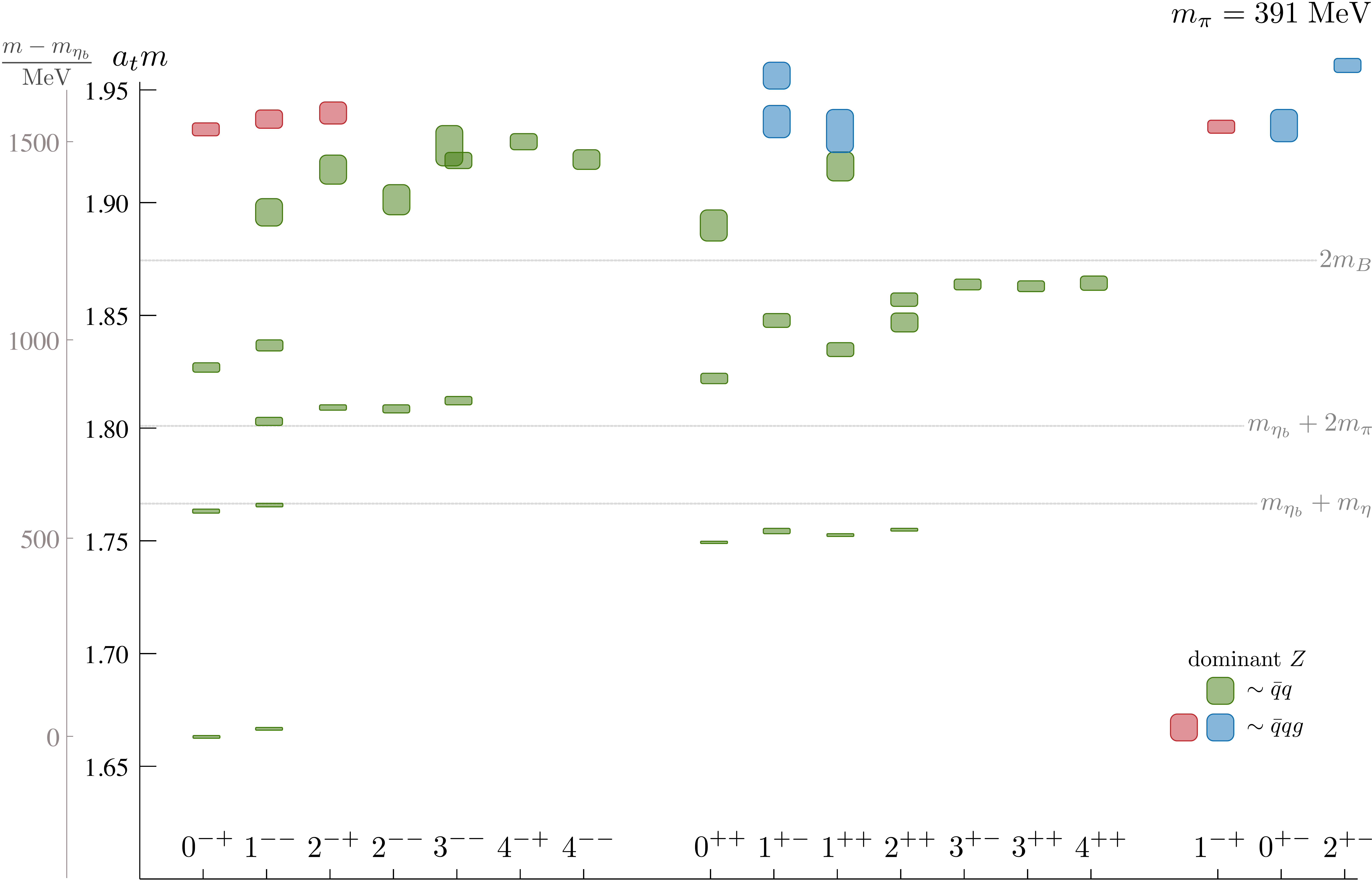}
  \caption{The spin-identified spectrum of bottomonium mesons, labelled by $J^{PC}$. Results are from the $20^3$ volume with two time sources, the highest-statistics ensemble in this study. The states are arranged by parity. The leftmost group are the negative parity mesons, middle group shows the positive parity mesons and the rightmost group shows the mesons with exotic quantum numbers. The vertical height of the boxes indicates the one-sigma statistical uncertainty. The grey dashed lines show the lowest-lying strong decay thresholds determined from this lattice data. States highlighted in red and blue are dominated by operators with a hybrid construction.}
  \label{fig:allJpc}
  \end{center}
\end{figure}

The spin-identified spectrum of bottomonium mesons labelled by continuum quantum numbers, $J^{PC}$, is shown in Figure~\ref{fig:allJpc}. 
The numerical values relative to the $\eta_b$, in MeV units, are provided in Table~\ref{tab:mass_by_JPC}. Using the operator overlaps to identify the different components of the given $J=2,3,4$ states split across lattice irreps, we do not find statistically significant discretisation effects due to rotational symmetry breaking, and therefore continuum spins of states up to $J=4$ are assigned. 

Clusters of states appear to follow quark model supermultiplets. For example, 
in the negative parity sector (leftmost on the plot) $(1,2,3)S$ and $(1,2)D$ are visible with possibly a few states of the $1G$ supermultiplet. We find no $0^{--}$ state in the energy region we consider. 
In the positive parity sector (middle pane) $(1,2)P$ and $1F$ are clearly identified. 
The left panel of Figure~\ref{fig:multiplet_overlaps} shows an example of the $1F$ operator overlaps for $\left( (\pi,\rho)\times D^{[3]}_{J_{13=2},\, J=3}\right)^J$, which are seen to be consistent across supermultiplet members. This enables us to identify that the third state in $2^{++}$ as $1F$, while the nearby second state forms part of the $2P$ supermultiplet.

The exotic quantum numbers, $1^{-+}$, $0^{+-}$ and $2^{+-}$ each contain a level, and there are additional levels not accounted for by the quark model supermultiplet counting in several irreps.

\subsection{Hybrid supermultiplets}
Candidate hybrid mesons are identified by examining the operator-state overlaps for each energy level in a given lattice
irrep. States with hybrid content are highlighted in red and blue in Fig.~\ref{fig:allJpc}, while all others appear in green. 

A state is proposed to be a hybrid if it is characterised by a relatively large overlap onto an operator proportional to the
field-strength tensor. The same procedure was followed in previous analyses of mesons in the light, strange, open-charm and
charmonium meson sectors. In each case a supermultiplet of hybrid states in $[(0,1,2)^{-+},1^{--}]$ was identified at an
energy scale $\approx 1.3$ GeV above the lightest state in the spectrum. It was previously commented that such a pattern
of states is predicted by bag models and models with constituent gluonic degrees of freedom, but appears to rule out some flux-tube
models. 

A similar pattern of hybrids is identified in this bottomonium system. The $T_1^{-+}$ irrep contains $J^{PC}=1^{-+}$, $3^{-+}$ and higher $J$, some of which are not exotic based on $q\bar{q}$ considerations. One difference to studies with lighter quarks is that a state dominated by a $J^{PC}=4^{-+}$ operator arises at a lower energy than the first level with $J^{PC}=1^{-+}$, as can be seen in Figures~\ref{fig:irrep_spectrum} and~\ref{fig:irrep_spectrum_L20_L24}. The second state in $T_1^{-+}$ is dominated by operators containing a field-strength tensor in their constructions, singling it out as a hybrid candidate. Furthermore, all of the irreps with $J^{PC}=(0,1,2)^{-+},1^{--}$ have a state dominated by such operators around $a_tm=1.93$. In $T_1^{--}$ this is state $\mathfrak{n}=8$, whose relative overlap was shown earlier in Figure~\ref{fig:skyscrapers}.

\begin{figure}[h]
\begin{center}
\includegraphics[width=0.99\textwidth]{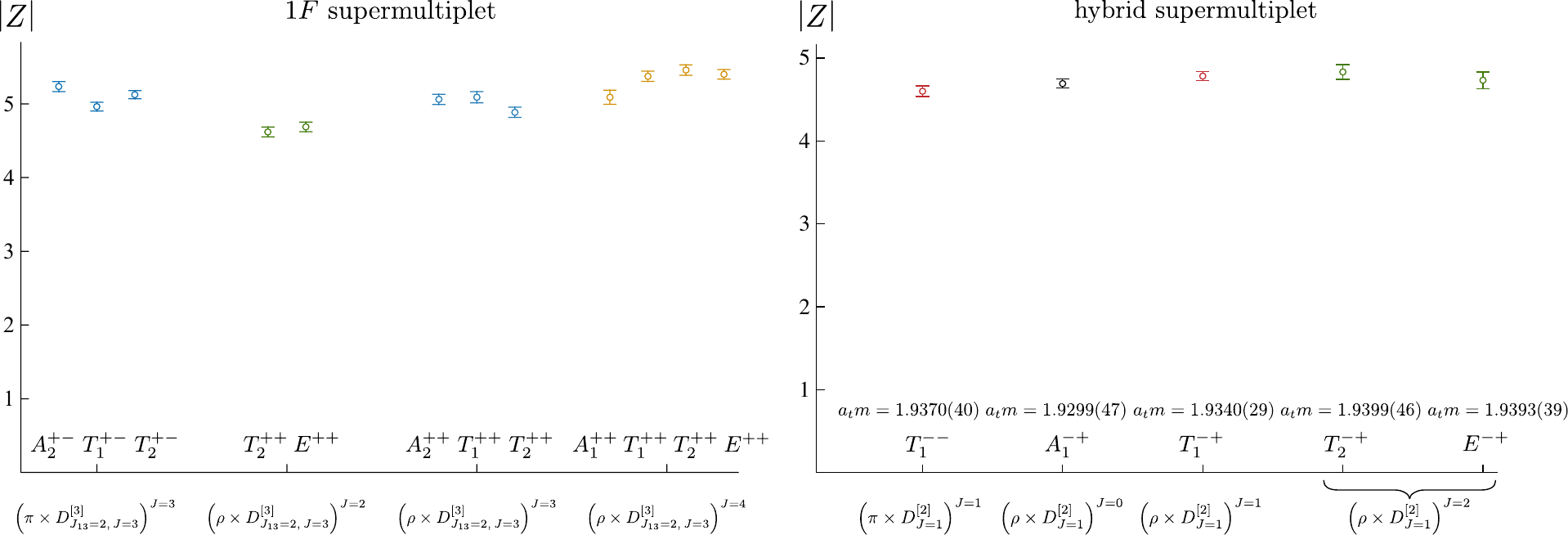}
\caption{Operator overlap values $|Z|$ for continuum operators corresponding to a $1F$ supermultiplet (left) and similarly a gluonic hybrid construction (right). These are subduced into different irreps as indicated, with a consistent value found for the operator overlap across irreps.}
\label{fig:multiplet_overlaps}
\end{center}
\end{figure}

In the right pane of Figure~\ref{fig:multiplet_overlaps}, we show the overlap of the simplest hybrid operator 
construction\footnote{We use the same notation as in Ref.~\cite{Liu:2012ze} where $\pi(\rho$) are $\gamma_5(\gamma_i$) 
respectively.}, $\left((\pi,\rho)\times D^{[2]}_{J=1}\right)^J$ and find that, as for 
the $1F$ supermultiplet, we can identify a hybrid supermultiplet across lattice irreps that has a common origin, with masses in the range $a_t m=1.93-1.94$. These states are highlighted in red in Fig.~\ref{fig:allJpc}. Other states identified as hybrid are highlighted in blue and may form parts of a higher-lying exotic supermultiplet.  In Figures~\ref{fig:irrep_spectrum} and~\ref{fig:irrep_spectrum_L20_L24}, the hybrid candidates have been highlighted with a dotted outline. This qualitative picture is very similar to that found with charm quarks in Refs.~\cite{Liu:2012ze} and~\cite{Cheung:2016bym}. Finally, we note that in 
Ref.~\cite{Brambilla:2019jfi} a similar pattern of bottomonium hybrids is 
inferred using the charmonium lattice calculations from Ref.~\cite{Liu:2012ze} as input, although many more states are predicted there than are found in this calculation.

\subsection{Comparison to experiment}
The aim of this study has been to investigate the qualitative features accessible in bottomonium using a relativistic action, and a basis of quark-antiquark operators in a proven methodology for extracting highly excited states. Although statistically precise, we do not necessarily expect these results to be quantitatively accurate for several reasons, including discretisation effects due to the heavy quark mass, the presence of additional levels due to hadronic decays, 
the larger-than-physical light quark mass and the lack of sea quark effects for bottom and charm. While 
a comprehensive error budget is beyond the scope of this work 
it is still informative to make a comparison with the existing experimental 
spectrum.

\begin{figure}[h]
  \begin{center}
    \includegraphics[width=0.95\textwidth]{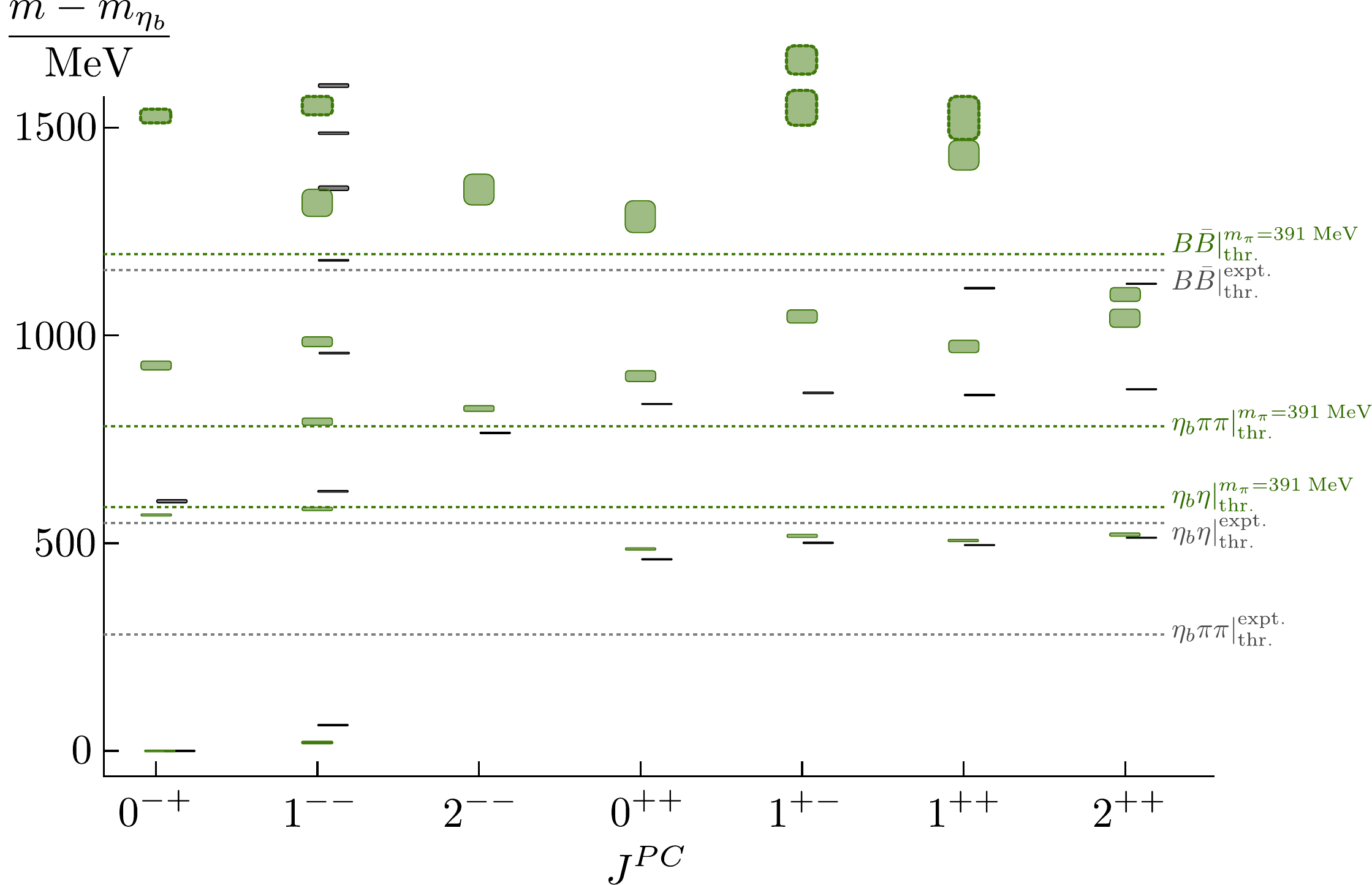}
  \caption{A selection of energy levels determined in this study for which there are corresponding experimental values. The energy levels obtained in this study, shown in green, compared with experiment~\cite{Tanabashi:2018oca}, shown in black.}
  \label{fig:lat_v_exp}
  \end{center}
\end{figure}

In Figure~\ref{fig:lat_v_exp}, we show the experimental energies listed in 
Ref.~\cite{Tanabashi:2018oca} compared with those determined in this work. 
The hyperfine splitting is smaller in the lattice spectra, compared with experiment, as discussed above.  
The $2S$ states lie slightly below their experimental counterparts.  
The $1P$ states are in reasonable agreement with experiment while the $2P$ multiplet is consistently higher in energy. 
Comparing spin-averaged splittings, we find that $\Delta_{2S-1S}=564(3)$~MeV is 
in good agreement with the corresponding experimental value of 563.3 MeV. The spin-averaged (1P-1S) splitting is 
$495(3)$ MeV in this study compared to 455.12 MeV from experiment.

An extra level belonging to the $1D$ supermultiplet appears in $1^{--}$, similar to those seen in quark potential models~\cite{Godfrey:1985xj}. This state has not been unambiguously observed experimentally although the $2^{--}$ member is known~\cite{Bonvicini:2004yj,delAmoSanchez:2010kz}. The comparable state in charmonium, the $\psi(3770)$, is clearly seen in both hadronic and $e^+e^-$ processes, although the lighter quark mass and the proximity of $D\bar{D}$ are key differences that may mix the $S$ and $D$-wave eigenstates and enlarge production rates. 

Quark potential models typically predict a small electromagnetic coupling to the $\Upsilon(1D)$~\cite{Kwong:1988ae} and 
it can in-principle be computed on the lattice from a local vector current of the form $\left<0|\bar{\psi}\gamma_i\psi|\Upsilon(1D)\right>$, for example as in Ref.~\cite{Shultz:2015pfa}. While such a calculation is beyond the scope of the 
present work, we already have the equivalent smeared vector operator\footnote{Otherwise referred to as $(\rho)^{J=1}$ following the operator naming scheme in Ref.~\cite{Dudek:2010wm}.}, visible as the fourth $|\tilde{Z}|$ plotted from the eigenvectors of each state in Fig.~\ref{fig:skyscrapers}. The $\Upsilon(1D)$ is identified as state $\mathfrak{n}=2$ where this vector operator makes a tiny contribution, unlike the majority of states in the spectrum identified as having $J^{PC}=1^{--}$. The pattern of overlaps suggests that state $\mathfrak{n}=5$ is the $\Upsilon(2D)$, this receives a slightly larger contribution from this smeared vector operator. We note that the state $\mathfrak{n}=8$, identified as a hybrid, also has a relatively small overlap with this smeared vector operator.

The mixing of the $S$-$D$ eigenstates in $1^{--}$ can be qualitatively assessed, once again from the operator overlaps. The final two non-hybrid vector operators shown in red in Fig.~\ref{fig:skyscrapers} ($\rho\times D^{[2]}_{J=2})^{J=1}$ and $\rho_2\times D^{[2]}_{J=2})^{J=1}$) have previously been found to have dominant overlaps onto $^{2S+1}L_J= {^{3}D_1}$ states~\cite{Dudek:2011bn}. The same is true here with a tiny overlap in states $\mathfrak{n}=0,1$ and a dominant overlap in state $\mathfrak{n}=2$. This, and the previous comments regarding the vector operator, suggests very weak $S$-$D$ mixing in the lowest three $1^{--}$ states\footnote{Corresponding to $\Upsilon(1S)$, $\Upsilon(2S)$ and $\Upsilon(1D)$ respectively.}. The same cannot be said of states $\mathfrak{n}=4,5$\footnote{Corresponding to $\Upsilon(3S)$ and $\Upsilon(2D)$ respectively.} where significant contributions from both operators are present. These states are found either side of $B\bar{B}$ threshold, not dissimilar to the $\psi(3770)$ and $\psi(2S)$ in charmonium found above and below $D\overline{D}$.

The operator basis used in this $T_1^{--}$ irrep contains 26 operators, many more than the 10 states we quote. Further levels are found at higher energies, but none lower in mass. We stress that this does not necessarily mean that further levels are not present, we know there must be hadron-hadron levels which are absent, and there is no guarantee the operators used will overlap well with all the physical eigenstates. Of the 26 operators used, 5 pairs are identical in the non-relativistic limit, which is a good approximation in bottomonium. The absence of any states belonging to the $4S$ and $5S$ supermultiplets is striking, but this should not be taken as evidence that they are not present, it is possible such states could be found in a more sophisticated calculation, even below the hybrids in $0^{-+}$ and $1^{--}$.

Similar differences can be found in $2^{++}$. Qualitatively, 3 states are found experimentally and 3 are present in this calculation, however, as discussed above and shown in the left panel of Fig.~\ref{fig:multiplet_overlaps}, the third level appears to be part of the $1F$ supermultiplet while the experimental candidate is assigned to $3P$. No clear $3P$ states arise in our calculation, but again, this absence should not be taken as clear evidence that they are not present in reality.

\section{Summary}
\label{sec:summary}

A first study of bottomonium including exotic and hybrid mesons from lattice QCD has been presented. 
An extensive pattern of states comparable to that predicted by quark models is determined. In addition, 
several hybrid meson candidates are identified at an energy approximately 1500 MeV above the ground state $\eta_b$, 
some of which can be grouped into a hybrid supermultiplet with $J^{PC}=(0,1,2)^{--},1^{-+}$. 
Similar qualitative features have previously been observed in studies spanning light, strange and charm quarks and 
now for the first time in bottom-quark physics.

This study was carried out with $2+1$ dynamical quark flavours using distillation for quark propagation.  
A single lattice spacing and a $b$-quark mass parameter slightly less than unity are used 
so that discretisation uncertainties remain unquantified. Within the 
scope of the calculation we have assessed their impact by considering dispersion relations of heavy-heavy and 
heavy-light pseudoscalar and vector mesons and by varying the 
coefficient of the spatial clover term $c_s$ in the heavy-quark action.
Rotation breaking effects are also found to be small and continuum spins are reliably assigned for states up to $J=4$. 
A large basis of carefully-constructed single-meson operators is used however hadron-hadron or multi-hadron operators 
are not included. These are expected to be relevant above the indicated hadron-hadron thresholds 
where additional eigenstates should be present that would mix with states extracted here. 
Several studies have demonstrated that the methodology used here produces 
an eigenstate within the hadronic width of the resonance in the vast majority of cases. Such resonance effects are 
highly volume dependent and we have tested that the same qualitative picture holds on a larger volume where in the 
majority of cases good quantitative agreement is found.

A similar determination of the $B$ meson spectrum and the $B_c$ spectrum is underway, and 
some preliminary results were presented in Ref.~\cite{Ryan:2020spd}. 
These methods appear to be sufficiently reliable that investigations of near-threshold states can now begin, 
presenting many interesting possibilities.

\bigskip

\begin{acknowledgments}
We thank our colleagues in the Hadron Spectrum Collaboration, in particular Jozef Dudek and Christopher Thomas 
for useful comments. 
DJW acknowledges support from a Royal Society University Research Fellowship and from 
the U.K. Science and Technology Facilities Council (STFC) [grant number ST/P000681/1. 
SR acknowledges the hospitality of the HEP group at DAMTP where this project was 
started and the support and hospitality of the Technical University of Munich - Institute for Advanced Study 
through the award of an August Wilhelm Scheer Visiting Professorship.
This work was performed using resources provided by the Cambridge Service for Data Driven Discovery (CSD3) operated by the University of Cambridge Research Computing Service (www.csd3.cam.ac.uk), provided by Dell EMC and Intel using Tier-2 funding from the Engineering and Physical Sciences Research Council (capital grant EP/P020259/1), and DiRAC funding from the Science and Technology Facilities Council (www.dirac.ac.uk). The DiRAC component of CSD3 was funded by BEIS capital funding via STFC capital grants ST/P002307/1 and ST/R002452/1 and STFC operations grant ST/R00689X/1. DiRAC is part of the National e-Infrastructure. 
Computations were also performed at Jefferson Laboratory under the USQCD Initiative and the LQCD ARRA project and on the Seagull cluster maintained by the Trinity Centre for High Performance Computing (TCHPC).

The software codes {\tt Chroma}~\cite{Edwards:2004sx}, {\tt QUDA}~\cite{Clark:2009wm,Babich:2010mu}, {\tt QPhiX}~\cite{ISC13Phi}, and {\tt QOPQDP}~\cite{Osborn:2010mb,Babich:2010qb} were used to compute the propagators required for this project.
This research was supported in part under an ALCC award, and used resources of the Oak Ridge Leadership Computing Facility at the Oak Ridge National Laboratory, which is supported by the Office of Science of the U.S. Department of Energy under Contract No. DE-AC05-00OR22725. This research is also part of the Blue Waters sustained-petascale computing project, which is supported by the National Science Foundation (awards OCI-0725070 and ACI-1238993) and the state of Illinois. Blue Waters is a joint effort of the University of Illinois at Urbana-Champaign and its National Center for Supercomputing Applications. This work is also part of the PRAC ``Lattice QCD on Blue Waters''. This research used resources of the National Energy Research Scientific Computing Center (NERSC), a DOE Office of Science User Facility supported by the Office of Science of the U.S. Department of Energy under Contract No. DEAC02-05CH11231. The authors acknowledge the Texas Advanced Computing Center (TACC) at The University of Texas at Austin for providing HPC resources that have contributed to the research results reported within this paper.

Gauge configurations were generated using resources awarded from the U.S. Department of Energy INCITE program at the Oak Ridge Leadership Computing Facility, the NERSC, the NSF Teragrid at the TACC and the Pittsburgh Supercomputer Center, as well as at Jefferson Lab.
\end{acknowledgments}

\appendix
\section{Masses summary table}

In table~\ref{tab:mass_by_JPC} we summarise the masses found in physical units relative to the $\eta_b$ as measured in this study, as presented above in Figure~\ref{fig:allJpc}.

\begin{table}[!h]
\begin{center}
\begin{tabular}{c|cccccc}
$J^{PC}$ & \multicolumn{3}{l}{$(m-m_{\eta_b})$/MeV} \\
\hline
$0^{-+}$ & 0.0(3)    & 568(2)   & 928(10) & 1527(16)$^\dagger$\\
$1^{--}$ & 20.1(4) & 582(4)   & 792(9) &  985(12) & 1319(33) & 1553(22)$^\dagger$\\
$2^{-+}$ & 827(7)  & 1426(36) & 1568(27)$^\dagger$ \\
$2^{--}$ & 824(7)  & 1351(37) \\
$3^{--}$ & 845(7)  & 1448(19) & 1487(50) \\
$4^{-+}$ & 1496(20)\\
$4^{--}$ & 1451(23)\\
\hline
$0^{++}$ & 486(2) & 902(13) & 1285(39)\\
$1^{+-}$ & 518(4) & 1045(15) & 1547(40)$^\dagger$ & 1663(33)$^\dagger$\\
$1^{++}$ & 506(3) & 973(15)  & 1434(35)           & 1522(53)$^\dagger$\\
$2^{++}$ & 521(4) & 1099(17) & 1042(21)\\
$3^{+-}$ & 1137(12)\\
$3^{++}$ & 1133(12)\\
$4^{++}$ & 1141(15)\\
\hline
$1^{-+}$ & 1536(16)$^\dagger$\\
$0^{+-}$ & 1536(39)$^\dagger$\\
$2^{+-}$ & 1689(17)$^\dagger$\\
\end{tabular}
\caption{Masses by $J^{PC}$ using $a_tm_{\eta_b}=1.66300(5)$ and $a_t^{-1}=5667$~MeV. $^\dagger$A level identified as a hybrid by its dominant operator overlaps.}
\label{tab:mass_by_JPC}
\end{center}
\end{table}

\bibliography{bottom_840_paper}
\bibliographystyle{JHEP}

\end{document}